\documentclass[journal]{IEEEtran}
\usepackage{amsmath,amsfonts}
\usepackage{algorithmic}
\usepackage{algorithm}
\usepackage{array}
\usepackage[caption=false,font=normalsize,labelfont=sf,textfont=sf]{subfig}
\usepackage{textcomp}
\usepackage{stfloats}
\usepackage{url}
\usepackage{verbatim}
\usepackage{graphicx}
\usepackage{cite}
\usepackage{color}
\begin{document}

\title{Channel Capacity Saturation Point and Beamforming Acceleration for Near-Field XL-MIMO Multiuser Communications}

\author{Xiangyu Cui, 
Ki-Hong Park, \textit{Senior Member, IEEE},  
and Mohamed-Slim Alouini, \textit{Fellow, IEEE}
\thanks{The authors are with the CEMSE Division, King Abdullah University of Science and Technology, Thuwal 6900, Makkah Province, Saudi Arabia (e-mail: xiangyu.cui@kaust.edu.sa, kihong.park@kaust.edu.sa, slim.alouini@kaust.edu.sa).}}

%\markboth{Journal of \LaTeX\ Class Files,~Vol.~xx, No.~x, August~2021}%
%{Shell \MakeLowercase{\textit{et al.}}: A capacity saturation limit for near-field XL-MIMO}


\maketitle

\begin{abstract}
One of the most important technologies in the fifth generation (5G) and the sixth generation (6G) is massive multiple input multiple outputs (MIMO) or extremely large-scale MIMO (XL-MIMO). With the evolving high-frequency technologies in millimeter band or tereHz band, the communication scene is changing into near-field rather than the conventional far-field scenario. In this letter, instead of advertising the XL-MIMO in the near-field, we appeal that a limit should be set on the size of the antenna array, beyond which the channel capacity will not show a significant increase. We show capacity saturation point can be analytically determined. Moreover, we propose a new beamforming algorithm that relieve the heavy computation due to the large antenna size even around the saturation point. Numerical results are provided to validate our analysis and show the performance of our newly proposed beamforming scheme.
\end{abstract}


\section{Introduction}
A fundamental technology driving the evolution of the fifth generation (5G) mobile communication system is massive multiple input multiple outputs (MIMO). Following the trend of enlarging the antenna array, the concept of extremely large-scale MIMO (XL-MIMO) is proposed for the future sixth-generation (6G) mobile communication system. Moreover, another trend is to use higher and higher frequency spectrum bands such as millimeter waves and terahertz waves for its abundant spectrum resources. As a result of these two trends above, users in future communication systems have a higher probability of situating in a near-field scenario, where the previous analysis for the far-field becomes inaccurate and outdated. Hence, it is urgent for us to renew our previous analysis into the new ones based on more accurate models.

There has already been some research done on the near-field communication. Ref. \cite{cui2024near} summarizes the previous power formulas in \cite{lu2021communicating,lu2021does} and derives a closed-form user correlation formula using the stationary phase method without any approximation on channel models, which provides intuitive insights on the underlying factors influencing the correlation. A notable discovery in \cite{cui2024near}, differing from previous research, is that the correlation between users is proven to converge to a constant rather than zero as the number of antennas approaches infinity. In \cite{lu2021near}, Lu and Zeng investigated the performance analysis for different beamforming schemes including zero-forcing (ZF), maximal ratio combining (MRC), and minimum mean square error (MMSE), based on which they propose that the near-field model can enable a new degree-of-freedom by distance separation. In \cite{zhao2025channel}, for different multiuser channel models, the performance analysis including the closed-form formula for capacity region is given. {In \cite{zhu2024mimo}, Zhu et al. prove the ergodic capacity saturation by electromagnetic information theory and find the corresponding saturation point for array-to-array system where a limit exists for the total array power, which is different from our multi-user scheme where the power limit is set on each individual user.} In \cite{xie2023performance}, Xie et al. derive a closed-form ergodic capacity with random user positions and show that the discrete MIMO performance will converge to the continuous one. In \cite{cui2022channel}, the authors give the channel vector coherence by the Fresnel approximation based on angular sampling and distant sampling, which is further applied to obtain sparsity in the polar domain. In \cite{lu2023near}, Lu and Dai propose a channel estimation method for the mixed line-of-sight and none-line-of-sight case. 

In most of the works mentioned above, it can be found the capacity of the near-field XL-MIMO system should converge to a limit instead of increasing unlimitedly as expected in the far-field model. As a result, to optimize cost efficiency, it is crucial to identify the point beyond which the capacity no longer increases. In this paper, based on our previous work \cite{cui2024near}, we analyze the antenna size required by the capacity saturation. Furthermore, we derive the ergodic saturation point and verify it by numerical simulation. To resolve the high calculation load caused by the large size of the saturation antenna array, we propose a new beamforming calculation method, which substitutes the large-scale inner product operation by the formulas in \cite{cui2024near}. The numerical verification shows that our modified method can reach the performance of the original beamforming at and beyond the saturation point.

\section{System Model}
We consider a narrowband multiuser uplink system with one base station equipped with an extremely large-scale antenna array receiving signals from single-antenna users which can be represented as
\begin{equation}
    \mathbf{y=Hx+n}
\end{equation}
where $\mathbf{H}$ is the channel matrix between the users and the antenna elements, $\mathbf{x}$ is the signal from different users with $\mathbb{E}[\mathbf{xx}^H] = P\mathbf{I}$ corresponding to the user transmitting power P, and $\mathbf{n}$ is the additive Gaussian noise (AWGN) at the receiving elements with the noise power $\sigma_n^2$.

Note that the downlink conclusions can be got in a similar way to this article. We omit it here for brevity. \footnote{The conclusions for uniform rectangular array and multipath occasions can be obtained in a similar way to this article. We may address these problems in future work.}

\subsection{Antenna Scheme}
In this paper, we consider a uniform linear array (ULA) equipped with $M$ antenna elements at the base station.

\begin{figure}[!t]
    \centering
    \subfloat[]{\includegraphics[width=0.5\linewidth]{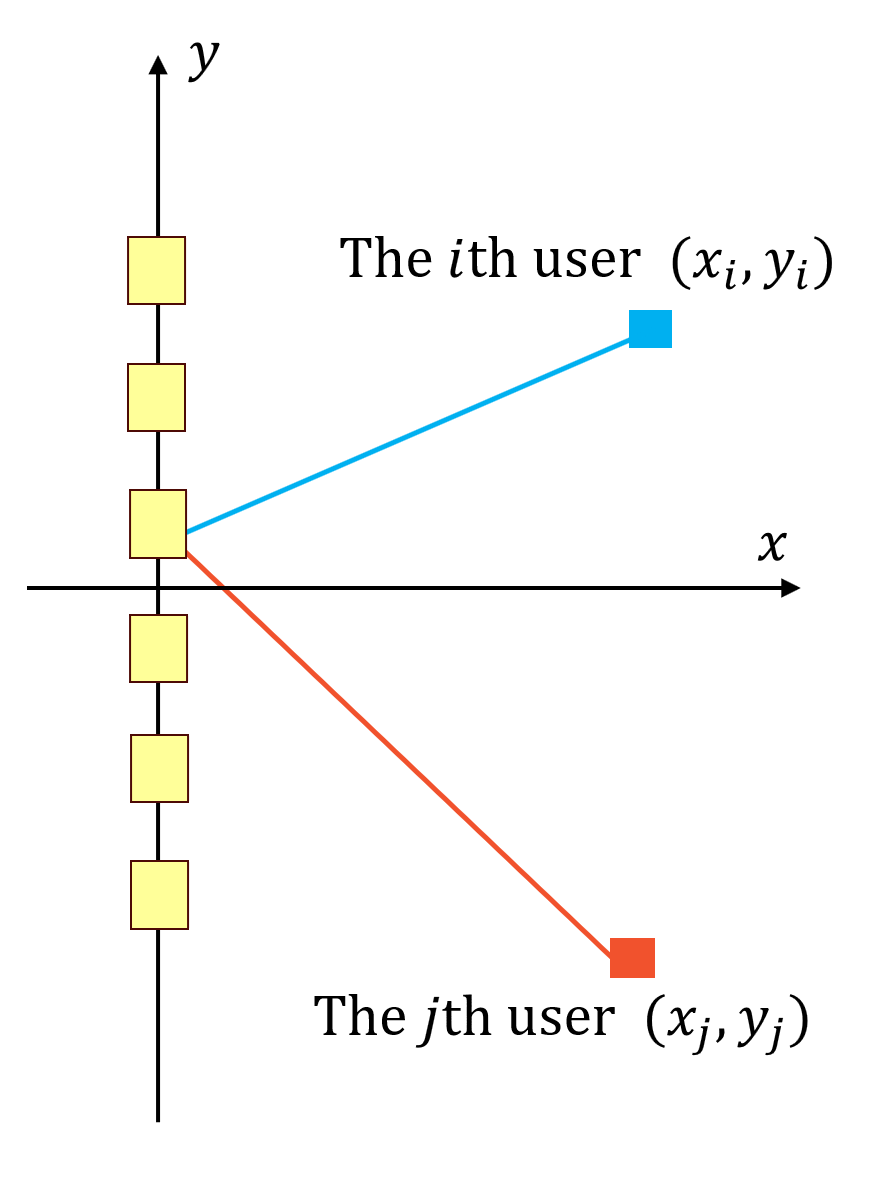}%
    \label{ULA_scheme}}
    \hfill
    \subfloat[]{\includegraphics[width=0.8\linewidth]{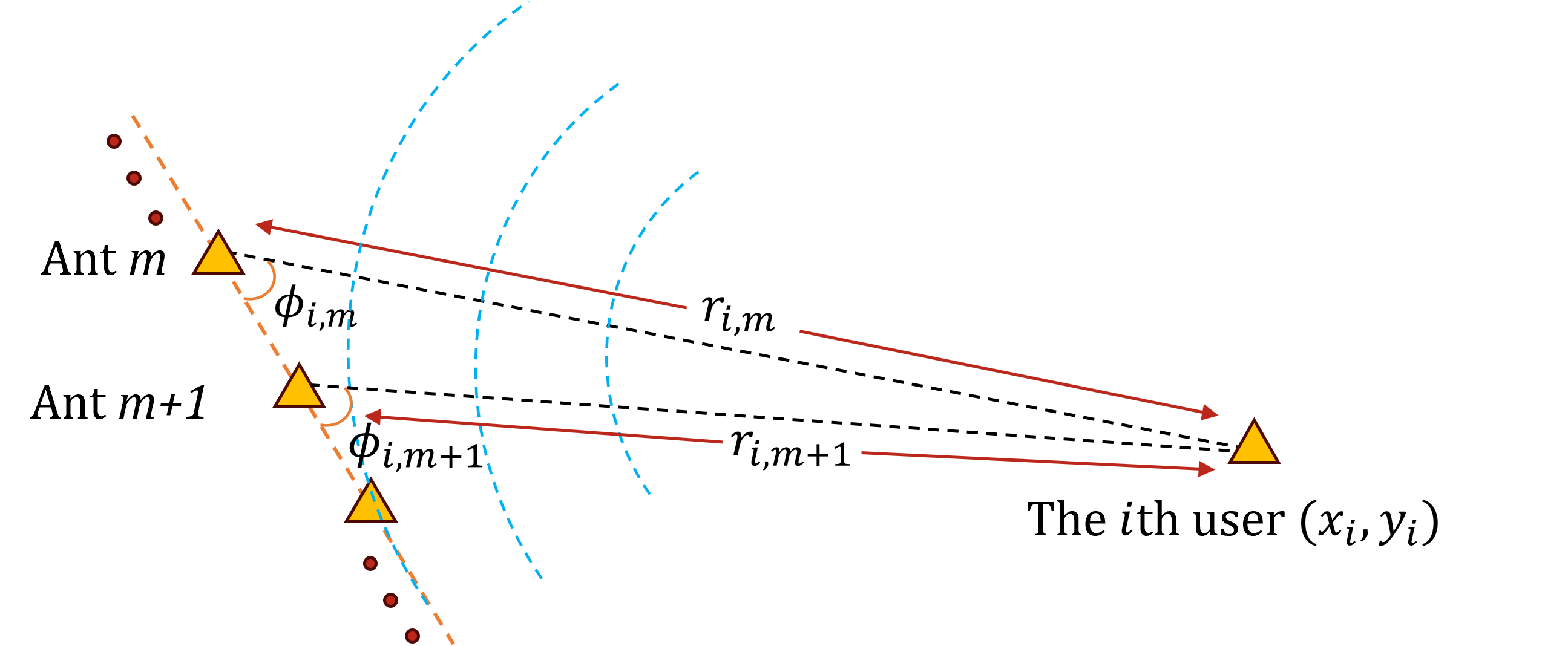}%
    \label{PNUSW_scheme}}
    \hfill
    \caption{(a) ULA scheme\cite{cui2024near} (b) PNUSW channel\cite{cui2024near}.}
\end{figure}
The antenna scheme is shown in Fig. \ref{ULA_scheme}.  These antenna elements in ULA are aligned in a line and two example users are indexed with $i$ and $j$ separately whose coordinates are $(x_i,y_i)$ and $(x_j,y_j)$. We set the adjacent distance between elements as half of the wavelength $\frac{\lambda}{2}$ to avoid correlation between elements.  Since a rotation of the user around the ULA make no change to the channel vector, we show the ULA and the users in 2-dimension. The ULA is set on the y-axis with its center at the origin. The channel vector for the $i$th user is expressed as
\begin{align}
    \mathbf{h}_i=[h_i(1),h_i(2),h_i(3),\dots,h_i(M)]^T.
    \label{single_path}
\end{align}
where $h_i(m)$ stands for the channel between the $m$th antenna element and the $i$th user.

\subsection{Channel Model}
In \cite{lu2021communicating}, the projected aperture is considered as an important factor of the channel and hence proposed the projected non-uniform spherical wave (PNUSW) model as shown in Fig. \ref{PNUSW_scheme}. 
To the best of the author's knowledge, one of the most accurate near-field channel models is the PNUSW model. Hence, we only consider the PNUSW model in our analysis. Note that only the line-of-sight channel is considered in this paper.

The channel between the $m$th antenna element and the $i$th user can be expressed as 
\begin{equation}
    \begin{aligned}
    h_i(m)=&\frac{\sqrt{\beta_0}e^{-j\frac{2\pi}{\lambda}r_{i,m}}}{r_{i,m}}\sqrt{\sin{\phi_{i,m}}}\\
    =& \frac{\sqrt{\beta_0}e^{-j\frac{2\pi}{\lambda}r_{i,m}}}{r_{i,m}}\sqrt{\frac{x_i}{r_{i,m}}}\\
    \end{aligned}
    \label{PNUSW_model_initial}
\end{equation}
where $\sqrt{\beta_0}$ serves as a constant to include the factors that may influence the received power such as antenna gain. Without loss of generality, we set $\beta_0$ as $1$ in this article. $r_{i,m}$ is the distance from the $i$th user to the $m$th antenna. $\phi_{i,m}$ is the incident angle from the $i$th user to the $m$th antenna. 

\section{Received Power and Correlation}
The received power for the $i$th user is defined as $P_i = P||\mathbf{h}_i||^2$ and the correlation between the $i$th user and the $j$th user is defined as $\rho_{ij}=\frac{\mathbf{h}_i^H\mathbf{h}_j}{||\mathbf{h}_i||||\mathbf{h}_j||}$.
In \cite{lu2021communicating}, a closed-form approximation for the received power is attained as
\begin{equation}
    \begin{aligned}
    P_i&\approx P\frac{2\beta_0}{\lambda x_i}\left(\frac{-4y_i+M\lambda}{\sqrt{16x_i^2+(M\lambda-4y_i)^2}}\right.\\
    &\left.+\frac{4y_i+M\lambda}{\sqrt{16x_i^2+(M\lambda+4y_i)^2}}\right)
    \label{power_for_ULA_PNUSW}
    \end{aligned}
\end{equation}
with a limit
\begin{equation}
    P_i^{\text{lim}}=\lim_{M\rightarrow\infty}P_i \approx P\frac{4\beta_0}{\lambda x_i}.
    \label{power_for_ULA_PNUSW_limit}
\end{equation}
According to \cite{cui2024near}, a closed-form approximation for the correlation by the new definition is attained by the stationary phase method (SPM) as
\begin{equation}
    \begin{aligned}
        \rho_{ij}\;&\approx\frac{\sqrt{\lambda} \left||x_i|-|x_j|\right|}{\left((|x_i|-|x_j|)^2+(y_i-y_j)^2\right)^\frac{3}{4}}\\
        &\times\!\left(\!\!\frac{-4y_i+M\lambda}{\sqrt{16x_i^2\!+\!(M\lambda\!-\!4y_i)^2}}\!+\!\frac{4y_i+M\lambda}{\sqrt{16x_i^2\!+\!(M\lambda\!+\!4y_i)^2}}\!\!\right)^{\!\!-\frac{1}{2}}\\
        &\times\!\left(\!\!\frac{-4y_j+M\lambda}{\sqrt{16x_j^2\!+\!(M\lambda\!-\!4y_j)^2}}\!+\!\frac{4y_j+M\lambda}{\sqrt{16x_j^2\!+\!(M\lambda\!+\!4y_j)^2}}\!\!\right)^{\!\!-\frac{1}{2}}\!\!\!\!\\
        &\times\!e^{j \text{sgn}(|x_i|-|x_j|)\left(\frac{2\pi}{\lambda}\sqrt{(|x_i|-|x_j|)^2+(y_i-y_j)^2}-\frac{\pi}{4}\right)}
    \end{aligned}
    \label{Correlation_PNUSW_ULA_exact}
\end{equation}
where
\begin{equation}
    \text{sgn}(x) =
    \begin{cases}
        -1 & \text{if } x \leq 0, \\
        1 & \text{if } x > 0.
    \end{cases}
\end{equation}
Without loss of generality, we can assume these two users are located on the same side of the ULA. Then we can simplify $\sqrt{(|x_i|-|x_j|)^2+(y_i-y_j)^2}$ as $d_{ij}$, the distance between users. As the number of antennas comes to infinity, the correlation between users come to a limit as
\begin{equation}
    \begin{aligned}
        \lim_{M\rightarrow\infty}\rho_{ij}
        &=\frac{\sqrt{\lambda}\left|\cos{\gamma_{ij}}\right|}{2\sqrt{d_{ij}}}e^{j \text{sgn}(|x_i|-|x_j|)\left(\frac{2\pi}{\lambda}d_{ij}-\frac{\pi}{4}\right)}
    \end{aligned}
    \label{Correlation_PNUSW_ULA_limit}
\end{equation}

\section{Capacity Saturation Point}
\label{Capacity_saturation}
According to the Shannon theorem, the channel capacity of multiuser uplink system is given by
\begin{equation}
    \begin{aligned}
        C = \log_2\det\left(\mathbf{I}+\frac{P}{\sigma_n^2}\mathbf{H}^H\mathbf{H}\right).
    \end{aligned}
\end{equation}

It has been shown that as the number of antenna elements increases the capacity will saturate to a limit in \cite{cui2024near,zhao2025channel}. It would be useful to get that saturation point, beyond which an increase in the number of antennas would not cause a significant increase in the capacity. Obviously, the saturation of capacity is caused by the saturation of matrix $\mathbf{H}^H\mathbf{H}$. 

It can be shown theoretically by our formulas in \cite{cui2024near} that the correlation between users are near zero with a small wavelength and a large distance between users. This indicates that the off-diagonal elements of $\mathbf{H}^H\mathbf{H}$ are nearly zero compared with the diagonal elements for a high communication frequency and a normal distance between users. Hence, the capacity saturation mainly relies on the saturation of the diagonal elements, which is verified in Appendix \ref{insig_off_diag}.

\subsection{Power Saturation Points}
\begin{figure}[!t]
    \centering
    \includegraphics[width=0.5\linewidth]{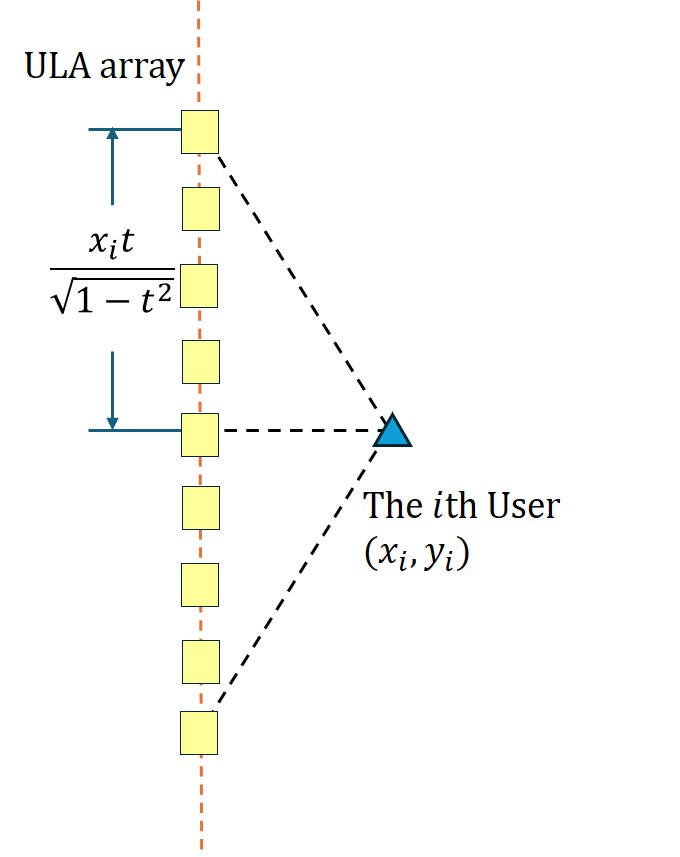}
    \caption{Power saturation point.}
    \label{power_saturation}
\end{figure}

The diagonal elements of $\mathbf{H}^H\mathbf{H}$ stand for the received power. Hence, we call the corresponding saturating points ``power saturation points". A reasonable definition of the saturation point is to set a threshold $t$ for the ratio between the exact power and the power limit:
\begin{equation}
\begin{aligned}
        \frac{P_i}{P_i^{\text{lim}}}\geq &t,\; t\leq 1\\
\end{aligned}
\end{equation}
which is
\begin{equation}
        \!\frac{-4y_i\!+\!M\lambda}{2\sqrt{16x_i^2\!+\!(M\lambda\!-\!4y_i)^2}}\!+\!\frac{4y_i+M\lambda}{2\sqrt{16x_i^2\!+\!(M\lambda\!+\!4y_i)^2}}\!\geq t.
\end{equation}
Without loss of generality, we assume $y_i=0$ at this moment, the corresponding solution would be
\begin{equation}
    M\geq \frac{4x_it}{\lambda \sqrt{1-t^2}}.
\end{equation}
The corresponding coordinates of the ULA elements on the $y$ axis are in the range of $\left[-\frac{x_it}{ \sqrt{1-t^2}},\frac{x_it}{ \sqrt{1-t^2}}\right]$.
as shown in Fig. \ref{power_saturation}.
Considering arbitrary $y_i$, we will get these two coordinates as
\begin{equation}
    \begin{aligned}
        y_{\text{up}}^i = \frac{x_it}{ \sqrt{1-t^2}}+y_i
    \end{aligned}
\end{equation}
and
\begin{equation}
    \begin{aligned}
        y_{\text{down}}^i = -\frac{x_it}{ \sqrt{1-t^2}}+y_i.
    \end{aligned}
\end{equation}
which is the saturation coordinates of a ULA antenna for a single user $(x_i,y_i)$ corresponding to the threshold $t$. 
With multiple users, the first capacity saturation coordinate would become
\begin{equation}
    y_{\text{up}} = \max_i{\frac{x_it}{ \sqrt{1-t^2}}+y_i}.
    \label{y_up}
\end{equation}
The other bound is defined as
\begin{equation}
    \begin{aligned}
        y_{\text{down}} = \min_i -\frac{x_it}{\sqrt{1-t^2}}+y_i.
    \end{aligned}
    \label{y_down}
\end{equation}

\subsection{Expectation of the saturation point}
Compared with the saturation point for a deterministic user, the ergodic saturation point is more representative of a user distribution. Here we give an example of a specific user distribution where $x_i \sim \mathcal{U}[x_{\text{min}}, x_{\text{max}}], 0<x_{\text{min}}<x_{\text{max}}$ and $y_i \sim \mathcal{U}[y_{\text{min}}, y_{\text{max}}],y_{\text{min}}<y_{\text{max}}$. In other words, the users are uniformed and distributed in a rectangle in the right side of the $y$ axis.

The expectation of (\ref{y_up}) is
\begin{equation}
    \begin{aligned}
        \mathbb{E}[y_{\text{up}}]=&R_2-\left(\frac{L_2 - 2 R_1 + R_2}{2(R_2-R_1)}\right)^K\frac{L_2 - 2 R_1 + R_2}{2(K+1)}\\
        &-\left(\frac{L_2 - 2 R_1 + R_2}{2(R_2-R_1)}\right)^K(R_1-L_1)\\
        &-\left(\frac{R_1-L_1}{2(R_2-R_1)}\right)^K\frac{R_1-L_1}{2(K+1)(2N+1)}\\
        &- \sqrt{2} K \sqrt{(R_1-L_1) (R_2-R_1)}\\
        &\times\mathrm{B}\left(\frac{R_1-L_1}{2 (R_2-R_1)}; \frac{3}{2}, K\right)
    \end{aligned}
    \label{up_expectation}
\end{equation}
where
\begin{eqnarray}
L_1 &=& \frac{x_{\text{min}}t}{\sqrt{1-t^2}}+y_{\text{min}}\nonumber \\
R_1 & =& \min(\frac{x_{\text{min}}t}{\sqrt{1-t^2}}+y_{\text{max}},\frac{x_{\text{max}}t}{\sqrt{1-t^2}}+y_{\text{min}}) \nonumber\\
L_2 &=& \max(\frac{x_{\text{min}}t}{\sqrt{1-t^2}}+y_{\text{max}},\frac{x_{\text{max}}t}{\sqrt{1-t^2}}+y_{\text{min}}) \nonumber\\
R_2 &=&\frac{x_{\text{max}}t}{\sqrt{1-t^2}}+y_{\text{max}} \nonumber    
\end{eqnarray}
and $K$ is the number of users. $\text{B}\left(\cdot;\cdot,\cdot\right)$ stands for the incomplete beta function. The proof is shown in Appendix \ref{ergodic_saturation_point}. The corresponding expectation of (\ref{y_down}) is
\begin{equation}
    \begin{aligned}
        \mathbb{E}[y_{\text{down}}]=y_{\text{min}}+y_{\text{max}}-\mathbb{E}[y_{\text{up}}].
    \end{aligned}
    \label{down_expectation}
\end{equation}
We show that the second to the fifth term in (\ref{up_expectation}) is monotonically increasing to a limit $0$ with the increase of $K$. The last term can also be shown to be converging to $0$ as $K$ comes to infinity, whose proof is shown in Appendix \ref{trend_beta}. As a result, we have the limits
\begin{equation}
    \begin{aligned}
        \lim_{K\rightarrow\infty} \mathbb{E}[y_{\text{up}}]=R_2=\frac{x_{\text{max}}t}{\sqrt{1-t^2}}+y_{\text{max}}
    \end{aligned}
    \label{lim_up}
\end{equation}
and
\begin{equation}
    \begin{aligned}
        \lim_{K\rightarrow\infty} \mathbb{E}[y_{\text{down}}]=y_{\text{min}}+y_{\text{max}}-R_2=y_{\text{min}}-\frac{x_{\text{max}}t}{\sqrt{1-t^2}}
    \end{aligned}
    \label{lim_down}
\end{equation}
which exactly corresponds to the saturation coordinates of the ``corner points'' of the uniform rectangular user distribution. In conclusion, as the number of users increases, the saturation points converge to those corresponding to the ones of ``corner points", which matches the intuition of ``short board effect".
\subsection{Numerical Verfication}
\label{verification_for_saturation}
\begin{figure}[!t]
    \centering
    \includegraphics[width=0.8\linewidth]{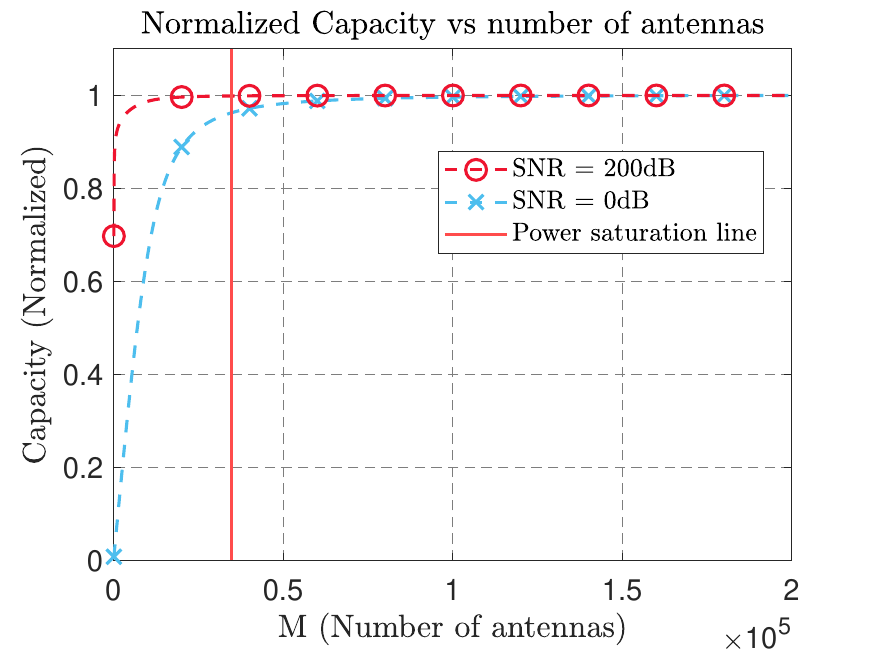}
    \caption{Normalized capacity vs the number of antennas.}
    \label{norm_cap}
\end{figure}
As in \cite{cui2024near}, we set the carrier frequency as $100$Ghz. The user distribution is $x_i\sim\mathcal{U}[1m,10m]$, $y_i\sim \mathcal{U}[-7.5m,7.5m]$ and $100$ users are assumed here for verification. In Fig. \ref{norm_cap}, to show that the saturation point is irrelevant to the signal-to-noise ratio (SNR), we set SNR as $0$dB and $200$dB for two different normalized capacity curves to verify our result at low and high SNR. The threshold $t$ is set as $0.9$ here. The red verticle line comes from the difference between (\ref{up_expectation}) and (\ref{down_expectation}), indicating the antenna size needed for the capacity saturation. 

As we expected, our closed-form formulas serve well as an estimation for the saturation point. Note that the convergence in high SNR is faster due to the logarithmic effect. In conclusion, the increase of the antenna elements beyond the saturation point will not cause a significant capacity increase. For cost-saving purposes, it is recommended to design a MIMO system not beyond $\left[-\frac{x_{\text{max}}t}{\sqrt{1-t^2}}+y_{\text{min}},\frac{x_{\text{max}}t}{\sqrt{1-t^2}}+y_{\text{max}}\right]$. 

\section{Beamforming Acceleration}
\subsection{Choosing ZF over MRC}
\label{ZF_over_MRC}
As in (\ref{Correlation_PNUSW_ULA_exact}), the correlation between users is a non-zero constant for the near-field model while the correlation based on uniform planar wave (UPW) model comes to zero as the number of antennas increases as in \cite{zhao2025channel, lu2021near}. Considering the fact that MRC can reach the channel capacity when the correlation between channels comes to zero, it is previously believed MRC should be used for the beamforming due to its lower complexity. 

However, in Fig. \ref{Capacity_vs_number_of_ant}, it is shown that ZF holds a higher capacity than MRC, reaching the channel capacity. It is because the founding assumption for UPW breaks down as the number of antennas increases, which makes the corresponding analysis no more accurate. As a result, though holding higher complexity, we should choose ZF (or MMSE) for a larger data rate.

\subsection{Inner product calculation speed-up}
As in part \ref{ZF_over_MRC}, algorithms like ZF provides a higher data rate while also coming with a higher computation complexity. One of the most time-consuming processes is the inner product calculation for elements in $\mathbf{H}^H\mathbf{H}$. 

With the fast development of integrated sensing and communication (ISAC), it has become possible to solve complex channel estimations with the assistance of sensing information\cite{liu2022integrated,you2025next}. Generally speaking, with the state of users including the distance, the angle, and the velocity, we may easily generate the channel and predict the future one, i.e. tracking. 

Based on the closed-form formulas (\ref{power_for_ULA_PNUSW}), (\ref{Correlation_PNUSW_ULA_exact}), and the techniques mentioned above, we can greatly simplify the calculation complexity of $\mathbf{H}^H\mathbf{H}$ and hence speed up the beamforming algorithms which require the calculation of $\mathbf{H}^H\mathbf{H}$. For instance, both the ZF and the linear minimum mean square error (MMSE) channel equalization require this calculation. For ZF, the equalization process can be represented as
\begin{equation}
    \begin{aligned}
    \hat{\mathbf{x}}=\left(\mathbf{H}^H\mathbf{H}\right)^{-1}\mathbf{H}^H\mathbf{y}
    \end{aligned}
\end{equation}
where the inner product calculation in $\mathbf{H}^H\mathbf{H}$ requires a high computational cost especially when the number of antennas becomes larger and larger, i.e. XL-MIMO. As an example, the $i$th diagonal element of $\mathbf{H}^H\mathbf{H}$ can be calculated as
\begin{equation}
    \begin{aligned}
        \mathbf{H}^H\mathbf{H}(i,i) \approx\frac{P_i}{P} 
    \end{aligned}
\end{equation}
where $P_i$ can be calculated by (\ref{power_for_ULA_PNUSW}) by simple system parameters and the user locations. The non-diagonal elements with the index $(i,j),i\neq j$ can be calculated by
\begin{equation}
    \begin{aligned}
        \mathbf{H}^H\mathbf{H}(i,j)\approx\frac{\rho_{ij}\sqrt{P_iP_j}}{P}
    \end{aligned}
\end{equation}
where $\rho_{ij}$ can be calculated by (\ref{Correlation_PNUSW_ULA_exact}) by simple system parameters and the user locations. We call this new method as ``modified ZF".
\subsection{Numerical Verification}
\begin{figure}[!t]
    \centering
    \includegraphics[width=0.8\linewidth]{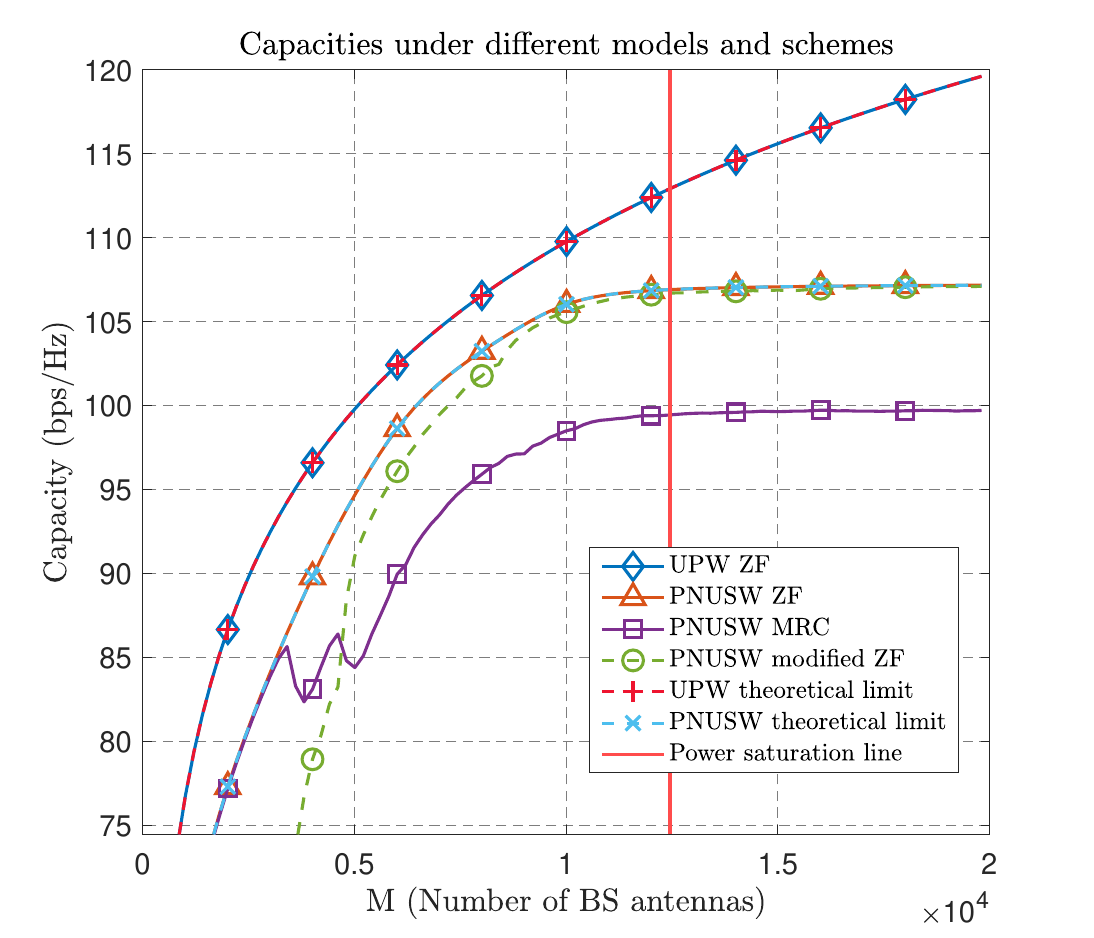}
    \caption{Capacity under different channel models and schemes}
    \label{Capacity_vs_number_of_ant}
\end{figure}

We use the same simulation settings in the section \ref{Capacity_saturation}. The only differences is that we change the user distribution to $x_i\sim \mathcal{U}(1m,2m)$ in the $x$-axis and we change the number of users to $10$. We also set the transmit power to suit a received SNR limit of around $33$dB for a single user at $(1.5m,0)$. Then we can get the capacity by using different channel models and beamforming schemes as shown in Fig. \ref{Capacity_vs_number_of_ant}.

As we can see in Fig. \ref{Capacity_vs_number_of_ant}, the modified ZF provides the same capacity as the original ZF after the saturation point, at which the inner product causes a large calculation complexity. For instance, the saturation point in Fig. \ref{Capacity_vs_number_of_ant} is around $10^4$ antennas which means $10^4$ multiplication operations and $10^4$ summation operations for one element in $\mathbf{H}^H\mathbf{H}$ are required in the original ZF. 

Note that the modified ZF has a worse performance than conventional ZF at the beginning since most of the elements in $\mathbf{H}^H\mathbf{H}$ will only converge after the corresponding saturation point. Hence, it is suggested to use the modified ZF for large-scale antenna systems to provide a higher date rate and lower computational complexity.

\section{Conclusion}
This paper proposes an antenna size limit for the near-field XL-MIMO multiuser system. A corresponding ergodic result is also given based on a uniform user distribution. It is found that the saturation size of the antenna array is decided by the boundary of the user distribution. Moreover, to cope with the heavy computation caused by the large saturation size, we accelerate the conventional ZF by substituting the inner product operation with the corresponding closed-form formulas. Simulation results are carried out to validate the ergodic saturation point and the performance of our modified ZF.

\appendices
\section{Insignificance of the off-diagonal terms in $\log_2\det{\left(\mathbf{I}+\frac{P}{\sigma_n^2}\mathbf{H}^H\mathbf{H}\right)}$}
\label{insig_off_diag}
For the capacity, we may compute the upper bound by using Hadamard's inequality as 
\begin{equation}
        U = \prod_{i=1}^K \left(1+a_i\right)\geq\det{\left(\mathbf{I}+\frac{P}{\sigma_n^2}\mathbf{H}^H\mathbf{H}\right)}
\end{equation}
where $a_i$ stands for the $i$th element of the diagonal elements of $\frac{P}{\sigma_n^2} \mathbf{H}^H\mathbf{H}$. Then we may find a lower bound by the Schur-Horn theorem and the Schur convexity as
\begin{equation}
    L = \det{\left(\mathbf{R}\right)}\prod_{i=1}^K (1+a_i)\leq \det{\left(\mathbf{I}+\frac{P}{\sigma_n^2}\mathbf{H}^H\mathbf{H}\right)}
    \label{lower_bound}
\end{equation}
where $\mathbf{R}$ is the correlation matrix whose elements are defined as $r_{ij}= \frac{\mathbf{h}_i^H\mathbf{h}_j}{||\mathbf{h}_i||||\mathbf{h}_j||}$. The matrix $\frac{P}{\sigma_n^2}\mathbf{H}^H\mathbf{H}$ can be decomposed as $\mathbf{P}^{\frac{1}{2}}\mathbf{R}\mathbf{P}^{\frac{1}{2}}$, where
\begin{equation}
    \begin{aligned}
        \mathbf{P} =\text{Diag}(a_1,a_2,\dots,a_K).
    \end{aligned}
\end{equation}
Hence,
\begin{equation}
    \begin{aligned}
        \det{\left(\frac{P}{P_n}\mathbf{H}^H\mathbf{H}\right)} =& \det{\left(\mathbf{P}\right)}\det{\left(\mathbf{R}\right)}\\
        =& \det{\left(\mathbf{R}\right)}\prod_{i=1}^K (1+a_i)\prod_{i=1}^K \frac{a_i}{1+a_i}\\
    \end{aligned}
\end{equation}
which means
\begin{equation}
\begin{aligned}
    \det{\left(\mathbf{R}\right)}\prod_{i=1}^K (1+a_i) =& \det{\left(\frac{P}{P_n}\mathbf{H}^H\mathbf{H}\right)}\prod_{i=1}^K \frac{1+a_i}{a_i}\\
    =& \prod_{i=1}^K \lambda_i\prod_{i=1}^K \frac{1+a_i}{a_i}.
\end{aligned}
\label{eq5}
\end{equation}
where $\lambda_i$ is the $i$th eigenvalue of $\frac{P}{P_n}\mathbf{H}^H\mathbf{H}$ with $\lambda_i>\lambda_j, i>j$. As a result, (\ref{lower_bound}) is equivalent to
\begin{align}
    \prod_{i=1}^K \left(1+\frac{1}{a_i}\right)\leq & \prod_{i=1}^K \left(1+\frac{1}{\lambda_i}\right)\\
    \sum_{i=1}^K \log_2\left(1+\frac{1}{a_i}\right)\leq &\sum_{i=1}^K \log_2\left(1+\frac{1}{\lambda_i}\right).\label{log_schur_convexity}
\end{align}
where (\ref{log_schur_convexity}) can be proved by the Schur-Horn theorem and the Schur convexity of the $\log_2(\cdot)$ function. The upper bound is fully determined by the diagonal elements while the lower bound is influenced by the off-diagonal elements. By showing that the gap between these two bounds is very small, we show that the non-diagonal elements have a minor influence on the final result. If we set the received SNR range as $[0\text{dB}, 30\text{dB}]$, the maximized normalized gap between them is
\begin{equation}
    \begin{aligned}
        \max_{\mathbf{a}}\frac{U-L}{U} &= \frac{-\log_2\det{(\mathbf{R})}}{\sum_{i=1}^K \log_2(1+1)}\\
        &=\frac{-K\frac{1}{K}\sum_{i=1}^K \log_2\lambda_i^R}{K}.\\
        & = -\frac{1}{K}\sum_{i=1}^K \log_2\lambda_i^R\\
        & = -\mathbb{E}_K \left[\log_2{\left(\lambda^R\right)}\right]
    \end{aligned}
    \label{eq9}
\end{equation}
which is the empirical expectation. $\lambda_i^R$ stands for the $i$th eigenvalue of the matrix $\mathbf{R}$. The off-diagonal elements of the correlation matrix are identically distributed but dependent, so the semicircle law does not apply. However, the eigenvalue distribution still converges as the number of users $K$ increases according to the simulation, and hence so does the empirical expectation.
\begin{figure}
    \centering
    \includegraphics[width=0.9\linewidth]{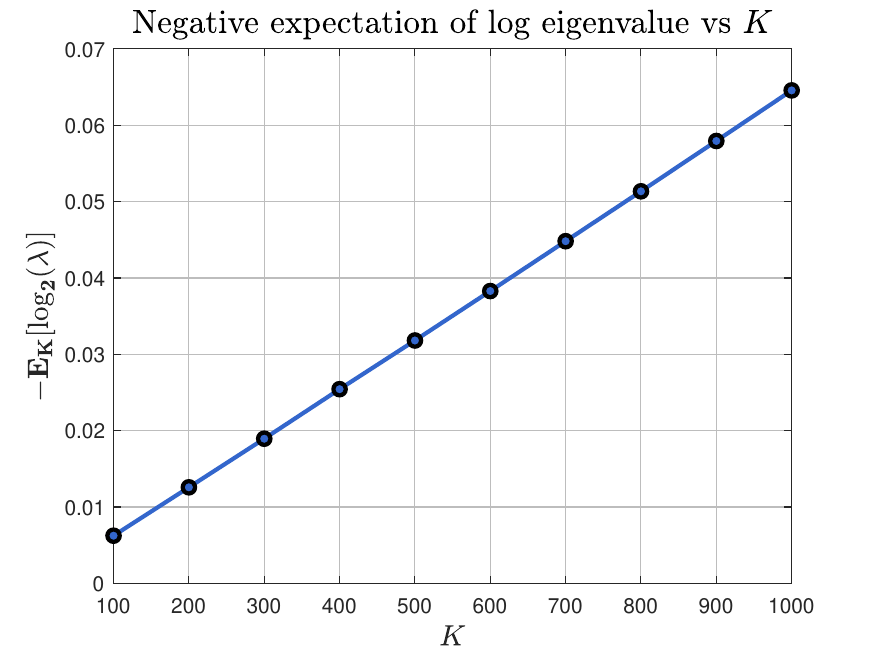}
    \caption{Negative empirial expectation vs the number of users $K$}
    \label{K_vs_E}
\end{figure}
In Fig. \ref{K_vs_E}, we show that the negative expectation increases as the number of users increases, where the simulation settings in part \ref{verification_for_saturation} are used. By our simulation above, we can make a conclusion that the inaccuracy increases with respect to the increase in the number of users $K$. The maximal inaccuracy is reached at $K=1000$ with the value as $0.064$ at a low receiving SNR at $0$dB. Note that the user distance between each other should not go beyond a certain limit for practicality. If we limit the distance between users beyond $d = 0.5m$, we would get the largest user density as $\frac{2}{\sqrt{3}d^2} \approx 4.619\;\text{users/}m^2$ by hexagonal packing, which sets a limit as $K\leq 15m\times 10m\times  4.619\; \text{users/}m^2\approx 693 \;\text{users}$. Hence, a user number further than $K=1000$ will not be considered due to the user density restriction. At last, a smaller inaccuracy can be achieved if we increase the receiving SNR assumption.

\section{Ergodic saturation point}
\label{ergodic_saturation_point}
It is easy to prove that the distribution of $z_i=\frac{x_it}{1-t^2}+y_i$ is
\begin{equation}
    f_{z_i}(z)=
    \begin{cases}
        &\frac{2}{D_1}\frac{1}{D_2}(z-L_1),z\in[L_1,R_1)\\ 
        &\frac{2}{D_1},z\in[R_1,L_2)\\
        &(R_2-z)\frac{2}{D_1}\frac{1}{D_2},z\in[L_2,R_2]\\
    \end{cases}
\end{equation}
where $D_1=L_2-R_1+R_2-L_1$ and $D_2=R_1-L_1$.
Hence, the cumulative distribution function (CDF) becomes 
\begin{equation}
    F_{z_i}(z) =
    \begin{cases}
        \frac{1}{D_1} \frac{(z - L_1)^2}{D_2}, & z \in [L_1, R_1), \\[10pt]
        \frac{(R_1 - L_1)}{D_1} + \frac{2}{D_1} (z - R_1), & z \in [R_1, L_2), \\[10pt]
        1 - \frac{1}{D_1} \frac{(z - R_2)^2}{D_2}, & z \in [L_2, R_2).
    \end{cases}
\end{equation}
Then we have the CDF of $y_{\text{up}}$
\begin{equation}
    \begin{aligned}
        F_{y_{\text{up}}}(y)=\left(F_{z_i}(y)\right)^K.
    \end{aligned}
\end{equation}
Accordingly, we have the its PDF as
\begin{equation}
    \begin{aligned}
        f_{y_{\text{up}}}(y) &= K\left( F_{z_i}(y)\right)^{K-1} f_{z_i}(y)
    \end{aligned}
\end{equation}
The desired expectation can obtained by
\begin{equation}
    \begin{aligned}
        \mathbb{E}[y_{\text{up}}]= \int_{L_1}^{R_2} f_{y_{\text{up}}}(y) y dy.
    \end{aligned}
\end{equation}
\section{The trend of incomplete beta function}
\label{trend_beta}
It is easy to prove that
\begin{equation}
    \begin{aligned}
        \text{B}(x;a,b)&=\int_0^x t^{a-1}(1-t)^{b-1}dt\\
        &\leq\int_0^1 t^{a-1}(1-t)^{b-1}dt\\
        &=\text{B}(a,b).
    \end{aligned}
\end{equation}
And we have
\begin{equation}
    \begin{aligned}
        \text{B}(a,b)=\frac{\Gamma{(a)}\Gamma{(b)}}{\Gamma{(a+b)}}
        \overset{(a)}{\sim}\Gamma(a)\frac{\Gamma(b)}{\Gamma(b)b^a}
        =\Gamma(a)b^{-a}
    \end{aligned}
\end{equation}
where (a) uses the asymptotical behavior of Gamma function as $b$ comes to infinity. By the deduction above, for our case, we have 
\begin{equation}
    \begin{aligned}
        &\sqrt{2} K \sqrt{(R_1-L_1) (R_2-R_1)}\mathrm{B}\left(\frac{R_1-L_1}{2 (R_2-R_1)}; \frac{3}{2}, K\right)\\
        <& \sqrt{2} K \sqrt{(R_1-L_1) (R_2-R_1)} \mathrm{B}\left(\frac{3}{2}, K\right)\\
        \sim& \sqrt{2} K \sqrt{(R_1-L_1) (R_2-R_1)} \Gamma\left(\frac{3}{2}\right) K^{-\frac{3}{2}}\\
        =& \sqrt{2}  \sqrt{(R_1-L_1) (R_2-R_1)} \Gamma\left(\frac{3}{2}\right) K^{-\frac{1}{2}}
    \end{aligned}
\end{equation}
which monotonically decreases to $0$ as the number of users comes to infinity.

\bibliographystyle{IEEEtran}
\bibliography{ref}

\end{document}